\documentclass[pre,aps,showpacs,byrevtex,amsmath,amssymb,twocolumn,preprintnumbers]{revtex4}

\usepackage{graphicx}% Include figure file

\begin{document}

\newcommand{\Cmm}{{C/m$^2$}}
\newcommand{\etal}{{\em et al.\/}}
\newcommand{\fj}{{}}

\title{A simple model for semipermeable membrane: Donnan equilibrium}

\author{Felipe Jim\'enez-\'Angeles}
\email{fangeles@imp.mx} \affiliation{Programa de Ingenier\'{\i}a
Molecular, Instituto Mexicano del Petr\'oleo, L\'azaro C\'ardenas
152, 07730 M\'exico, D. F., M\'exico}

\affiliation{Departamento de F\'{\i}sica, Universidad Aut\'onoma
Metropoloitana-Iztapalapa, Apartado Postal 55-334, 09340 D.F.
M\'exico}

%
%
%4th Author
\author{Marcelo Lozada-Cassou}
\email{marcelo@imp.mx} \affiliation{Programa de Ingenier\'{\i}a
Molecular, Instituto Mexicano del Petr\'oleo, L\'azaro C\'ardenas
152, 07730 M\'exico, D. F., M\'exico}

\affiliation{Departamento de F\'{\i}sica, Universidad Aut\'onoma
Metropoloitana-Iztapalapa, Apartado Postal 55-334, 09340 D.F.
M\'exico}
%

%
%
%2nd Author
%\author{Marcelo Lozada-Cassou}
%\email{marcelo@www.imp.mx} \affiliation{Programa de
%Ingenier\'{\i}a Molecular, Instituto Mexicano del Petr\'oleo,
%L\'azaro C\'ardenas 152, 07730 M\'exico, D. F., M\'exico}
%\affiliation{Departamento de F\'{\i}sica, Universidad Aut\'onoma
%Metropoloitana-Iztapalapa, Apartado Postal 55-334, 09340 D.F.
%M\'exico}

%
%

\begin{abstract}
We study a model for macroions in an electrolyte solution confined
by a semipermeable membrane. The membrane finite thickness is
considered and both membrane surfaces are uniformly charged. The
model explicitly includes electrostatic and size particles
correlations. Our study is focused on the adsorption of macroions
on the membrane surface and on the osmotic pressure. The
theoretical prediction for the osmotic pressure shows a good
agreement with experimental results.

\end{abstract}
%
%\pacs{87.15Aa, 36.20.Ey, 82.70.Dd}
% other packs , 61.20.Qg,
\maketitle

\section{Introduction}

Physics of two ionic solutions separated by a semipermeable
membrane is of wide interest in cell biology and colloids science
\cite{hoppe,himenz,tanford}.  A thermodynamical study of this
problem was first carried out by F. G. Donnan
\cite{donnan,guggenheim}, considering the two fluids phases (here
referred as $\alpha$ and $\beta$) in the following way: (i) the
$\alpha$-phase contains two (small) ionic species, (ii) the
$\beta$-phase contains the same ionic species as the
$\alpha$-phase plus one macroion species. The two fluid phases are
separated by a membrane which is permeable to the small ions and
impermeable to macroions, therefore, they interchange small ions
whereas macroions are restricted to  the $\beta$-phase. The
permeability condition is imposed {\fj by assuming (in both
phases)} a constant chemical potential of the permeating species.
In this simple model, Donnan derived an expression for the osmotic
pressure {\fj (}in terms of the ionic charge, concentration and
excluded volume{\fj )} which well describes systems close to
ideality. Historically, this problem has been known as Donnan
equilibrium.
%%
%%%%%%%%%%%%%%%%%%%%%%%%%%%%%%%%%%%%%%%%%%%%%%%
More recently some theories have been proposed to interpret
osmotic pressure data \cite{vincent80,youseff98}. These theories
consider {\fj phenomenologically the} macroion-macroion and
ion-macroion many-body interactions, {\fj and} provide a better
fit for the osmotic pressure of macroions solutions than Donnan
theory.

The surface of a biological membrane has a net charge when it is
in aqueous solution  \cite{hoppe} {\fj thus}, at a fluid-membrane
interface a broad variety of phenomena occur. It is known that
ionic solutions  in the neighborhood of a charged surface produce
an exponentially decaying charge distribution, known as the
electrical double layer. Low-concentrated solutions of monovalent
ions are well described by the Gouy-Chapman theory
(Poisson-Boltzmann equation) \cite{guy,chapman}. However,
multivalent ions display important deviations from this picture
\cite{rondelez_1998} {\fj and} more powerful theories from modern
statistical mechanics (such as molecular simulation
\cite{valleau80,lozada82,svensson_1989}, density functionals
\cite{percus64,evans,Percus_1990} and integral equations
\cite{henderson76,lozada81,plischke88,kjellander_1994,attard_1996})
{\fj have been implemented} for studying ions adsorption and
interfacial phenomena. Macroions adsorption is a subject of
current interest: In molecular engineering, the macroions
adsorption mechanisms are basic in self-assembling polyelectrolyte
layers on a charged substrate \cite{decher97} and novel colloids
stabilization mechanisms \cite{tohver_2001}.

{\fj By means of integral equations}, a previous study of Donnan
equilibrium has been carried out by Zhou and Stell
\cite{stell88_1,stell88_2}. They used the method proposed by
Henderson {\etal} \cite{henderson76}, {\fj which can be described
as follows}: starting from a semipermeable spherical cavity
\cite{note2}, the planar membrane is obtained taking the limit of
infinite cavity radius. Within this model, they obtained the
charge distribution and mean electrostatic potential. However, due
to the approximations used, they end up just with the integral
version of the linear Poisson-Boltzmann equation. A general
shortcoming of the Poisson-Boltzmann equation is that ionic size
effects (short range correlations) are completely neglected, in
consequence, the description of interfacial phenomena and
computation of thermodynamical properties is limited and valid
only for low values of charge and concentration.

From previous studies of two fluid phases separated by a {\em
permeable} membrane, it is known that the adsorption phenomena are
strongly influenced by the membrane thickness
\cite{lozadaPRL_1996,lozadaPRE_1997,aguilar2001}. On the other
hand, short range correlations influence effective colloid-colloid
interaction
\cite{muthuJCP_1996,crockerPRL_1999,likecharges1,likecharges2,likecharges4},
thermodynamical properties \cite{HansenPRL_2002} and adsorption
phenomena \cite{hendersonPRE_2001,jimenezPRL_2003} in colloidal
dispersions. From these antecedents it is seen that there are
several relevant aspects not considered in previous studies of
Donnan equilibrium which deserve a proper consideration. In this
study we consider explicitly the following effects: many-body
(short and long range) correlations, the membrane thickness and
the surface charge densities on each of the membrane faces. {\fj
We use {\em simple} model interactions and our study is carried
out by means of integral equations}. {\fj The} theory gives the
{\fj particles distribution in the neighborhood of} the membrane,
from {\fj which}, the osmotic pressure is calculated. There are
two points that we will address in this study: the adsorption of
macroions at the membrane surface and the computation of the
osmotic pressure for macroions solutions. Concerning the
adsorption phenomena, we observe a broad variety of phenomena:
charge reversal, {\fj charge inversion} and macroions adsorption
on a like-charged surface due to the fluid-fluid correlation. The
computed osmotic pressure is compared with experimental results
for a protein solution, obtaining an excellent agreement over a
wide regime of concentrations.

The paper is organized as follows: In section~\ref{theory} {\fj
we} describe the integral equations method {\fj and} the membrane
{\fj and} fluid models. {\fj In the same section, we derive} the
{\em hypernetted chain/mean spherical} (HNC/MS) integral equations
for the semipermeable membrane and the equations to compute the
osmotic pressure. In section \ref{results} a variety of results
are discussed and finally in section \ref{conclusions} some
conclusions are presented.

\section{Theory}
\label{theory}

\subsection{Integral equations for inhomogeneous fluids}
\label{theory:a} The method that we use to derive integral
equations for inhomogeneous fluids makes use of a simple fact: In
a fluid, an external field can be considered as a particle in the
fluid, i. e., as one more species infinitely dilute. This
statement is valid in general, however,  it is particularly useful
in the statistical mechanics theory for inhomogeneous fluids
\cite{lozada81,henderson92a} described below.

The multi-component Ornstein-Zernike equation for a fluid made up
of $n+1$ species is
%%
%
%%%%%%%%%%%%%%
\begin{equation}
 h_{i j}({\bf r}_{21}) \; =
    \; c_{i j}({\bf r}_{21}) + \sum_{m=1}^{n+1} \rho_m
     \int h_{i m}({\bf r}_{23}) c_{m j}({\bf r}_{13}) \,d{v}_{3},
 \label{ozh}
\end{equation}
%
%%%%%%%%%%%%%
where $ \rho_m$ is the number density of species $m$, $h_{i
j}({\bf r}_{21}) \equiv g_{ij}({\bf r}_{21})-1$ and $c_{i j}({\bf
r}_{21})$ are the total and direct correlation functions for two
particles at ${\bf r}_{2}$ and ${\bf r}_{1}$ of species $i$ and
$j$, respectively; with $g_{ij}({\bf r}_{21})$ the pair
distribution and ${\bf r}_{21} =  {\bf r}_{2}-{\bf r}_{1}$. Among
the most known closures between $h_{i j}({\bf r}_{21})$ and $c_{i
j}({\bf r}_{21})$ used to solve Eq.~(\ref{ozh}), {\fj we
have}\cite{hansen}

\begin{eqnarray}
  c_{i j}({\bf r}_{21})& = &-\beta u_{ij}({\bf r}_{21})+
  h_{i j}({\bf r}_{21})-\ln g_{ij}({\bf r}_{21}),
  \label{eq:chnc}\\
  c_{ij}({\bf r}_{21}) & = &  f_{ij}({\bf r}_{21})\exp\{-\beta u_{ij}({\bf r}_{21})\}g_{ij}({\bf
  r}_{21}),\\
 \label{eq:cpy}
  c_{ij}({\bf r}_{21}) & = & -\beta u_{ij}({\bf r}_{21})\;\; \mbox{for $r_{21}\equiv|{\bf r}_{21}|  \ge a_{ij}$}.
 \label{eq:cmsa}
\end{eqnarray}
%%%
%%%%
Eqs.~(\ref{eq:chnc}) to (\ref{eq:cmsa}) are known as the
hypernetted chain (HNC), the Percus-Yevick (PY) and the mean
spherical (MS) approximations, respectively; $u_{ij}({\bf
r}_{21})$ is the direct interaction potential between two
particles of species $i$ and $j$, $a_{ij}$ is their closest
approach distance, $f_{ij}({\bf r}_{21})\equiv \exp\{-\beta
u_{ij}({\bf r}_{21})\}-1$, and $\beta \equiv 1/k_B T$. Some more
possibilities to solve Eq.~(\ref{ozh}) are originated by
considering a closure for $c_{i j}({\bf r}_{21})$ in the first
term of Eq.~(\ref{ozh}) and a different one for $c_{mj}({\bf
r}_{13})$ in the second term of Eq.~(\ref{ozh}), giving rise to
hybrid closures.

To derive integral equations for inhomogeneous fluids, we let an
external field to be one of the fluid species, say $(n+1)$-species
(denoted as the $\gamma$-species), which is required to be
infinitely dilute, i. e., $\rho_\gamma \to 0$. Therefore, the
total correlation function between a $\gamma$-species particle and
a $j$-species particle is given by
\begin{eqnarray}
  h_{\gamma j}({\bf r}_{21}) \; =
    \; c_{\gamma j}({\bf r}_{21}) + \sum_{m=1}^{n} \rho_m
    \int h_{\gamma m}({\bf r}_{23}) c_{m j}({\bf r}_{13}) \,d{v}_{3}\nonumber \\
{\rm with} \quad j=1,...,n.
 \label{eq:oz}
\end{eqnarray}
The total correlation functions for the remaining species satisfy
a $n$-component Ornstein-Zernike equation as Eq.~(\ref{ozh}) (with
no $\gamma$ species) from which $c_{m j}({\bf r}_{13})$ is
obtained.
In this scheme, the pair correlation functions, $g_{\gamma i}({\bf
r}_{21})$, is just the inhomogeneous one-particle distribution
function, $g_{i}({\bf r}_1)$, for particles of species $j$ under
the influence of an external field. Thus, $h_{\gamma j}({\bf
r}_{21})$ and $c_{\gamma j}({\bf r}_{21})$ can be replaced with
$h_{j}({\bf r}_{1})\equiv g_{j}({\bf r}_{1})-1$ and $c_{j}({\bf
r}_{1})$, respectively. Thus, the inhomogeneous local
concentration for the $j$ species is given by
%%%%%%%%%%%%
%%%%%%%%%%%%
\begin{equation}
\rho_j({\bf r}_1) = \rho_j g_j({\bf r}_1), \label{eq:local_conc}
\end{equation}
By using the HNC closure (Eq.~(\ref{eq:hnc})) for $c_{\gamma
j}({\bf r}_{21})$ in Eq.~(\ref{eq:oz}), we get
%%%
%%%
\begin{equation}
  g_{j}({\bf r}_{1}) = \exp \left\{
  -\beta  u_{j}({\bf r}_{1}) + \sum_{m=1}^{n} \rho_m
     \int h_{m}({\bf r}_{3}) c_{m j}({\bf r}_{13}) \,d{v}_{3}
     \right\},
  \label{eq:hnc}
\end{equation}
where the subindex $\gamma$ has been omitted for consistency with
Eq.~(\ref{eq:local_conc}). {\fj In our approach, $c_{m j}({\bf
r}_{13})$ in the integral of Eq.~(\ref{eq:hnc}), is approximated
by the direct correlation function for a $n$-component homogeneous
fluid. Thus, $c_{m j}({\bf r}_{13})$ is obtained from
Eq.~(\ref{ozh}) using one of the closures provided by
Eqs.~(\ref{eq:chnc})-(\ref{eq:cmsa}). For the present derivation
we will use $c_{m j}({\bf r}_{13})$ obtained with the MS closere
(Eq.~(\ref{eq:cmsa})),} therefore, we obtain the hypernetted
chain/mean spherical (HNC/MS) integral equations for an
inhomogeneous fluid. This equation has shown to be particularly
successful in the case of inhomogeneous charged fluids when it is
compared with molecular simulation data
\cite{degreve93,lozadaPRE_1996,degreve97,deserno2001}.

\subsection{The semipermeable membrane and fluid models}

%%
%
%Fig1
%%%%%%%%%%%%%%%%%%%%%%%%%%%%
\begin{figure}
\includegraphics[width=8cm]{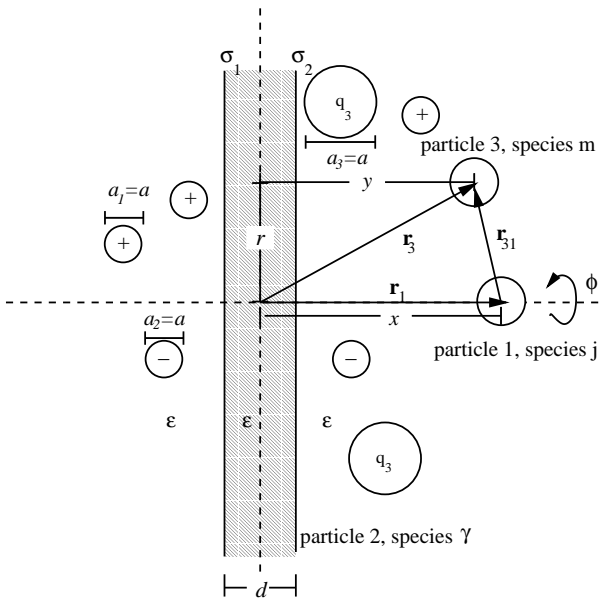}
\caption{Schematic representation for a model macroions solution
confined by a semipermeable membrane. For the integration of
Eq.~(\ref{eq:msa}) it is considered the {\em axial} symmetry
around the ${\bf r}_1$ vector where we fix a cylindrical
coordinates system ($\phi,r,y$).} \label{setup}
\end{figure}
%%%%%%%%%%%%%%%%%%%%%%

The membrane is modelled as a planar hard wall of thickness $d$,
and charge densities $\sigma_1$ and $\sigma_2$ on each surface and
separates two fluid phases, referred as $\alpha$ and $\beta$. In
Fig.~\ref{setup}, $\sigma_1$ and the $\alpha$-phase are at the
left hand side, whereas $\sigma_2$ and the $\beta$-phase are at
the right hand side. The fluid phased are made up in the following
way: (i) The $\alpha$-phase is a two component electrolyte,
considering the components to be hard spheres of diameter $a_i$
with a centered point charge $q_i=z_ie$ (being $z_i$ and $e$ the
ionic valence and the proton's charge, respectively) and $i=1,2$
standing for the species number. The solvent is considered as a
uniform medium of dielectric constant $\varepsilon$. (ii) The
$\beta$-phase is considered in the same way  as the $\alpha$-phase
(same solvent dielectric constant and containing the same ionic
species) plus one more species (species $3$) of diameter $a_3$ and
charge $q_3$. For simplicity, we have considered that the membrane
dielectric constant is equal to that of the solvent. In addition
it is considered that
\begin{equation}
\begin{array}{ll}
a &\equiv a_{1}=a_{2}  \leq a_{3},\\
z &\equiv z_{1}=-z_{2}.
\end{array}
\end{equation}
Hence, we refer to the third species as the macroion species. Two
ions of species $m$ and $j$ with relative position $r$, interact
via the following potential

\begin{equation}
  u_{mj}(r) \; = \; \left\{
    \begin{array}{ll}
      \infty   & \mbox{for $r <   a_{mj}$}, \\[2ex]
      {\displaystyle \frac{z_m z_j e^2}{\varepsilon r}}&
      \mbox{for $r \ge a_{mj}$},
    \end{array}
   \right.
 \label{ion_ion}
\end{equation}
with $m,j=1,\ldots,3$ and ${\displaystyle
a_{mj}\equiv\frac{a_{m}+a_{j}}{2}}$. Far away from the membrane
each phase is homogeneous and neutral, thus the neutrality
condition is written as
%%%
\begin{equation}
\sum_{i=1}^{2}z_{i}\rho_{i}^{\alpha}=\sum_{i=1}^{3}z_{i}\rho_{i}^{\beta}=0,
\label{neutrality}
\end{equation}
being $\rho_i^\alpha$ and $\rho_i^\beta$ the bulk concentrations
of the $i$ species in the $\alpha$ and $\beta$ phases,
respectively. The charge on the membrane is compensated by an
excess of charge in the fluid (per unit area), $\sigma'$
\cite{lozada84,lozada90b}:
\begin{equation}
 \sigma'\equiv\sigma^{\alpha}+\sigma^{\beta}=-\sigma_T,
  \label{sigma}
\end{equation}
with $\sigma_T=\sigma_1+\sigma_2$ and being $\sigma^{\alpha}$ and
$\sigma^{\beta}$ the excess of charge in the $\alpha$-phase and
$\beta$-phase, respectively, which are given by
\begin{equation}
\sigma^{\alpha}=\int_{-\infty}^{-\frac{d}{2}}\rho_{el}(x)d{x}
\label{sigma1}
\end{equation}
and
\begin{equation}
\sigma^{\beta}=\int^{\infty}_{\frac{d}{2}}\rho_{el}(x)d{x},
\label{sigma2}
\end{equation}
being
%%%%%
\begin{equation}
\rho_{el}(x) \equiv e \sum_{m=1}^{3} z_m \rho_{m}(x),
\label{sigma3}
\end{equation}
%%%%%%%%
%%%%%%%%
{\fj the local charge density profile}. In Eqs.~(\ref{sigma1}),
(\ref{sigma2}) and (\ref{sigma3}) it has been considered that the
particles reduced concentration profile ($g_{j}({\bf r}_{1})$)
depends only on the perpendicular position to the membrane, $x$,
i. e., $g_{j}({\bf r}_{1})=g_{j}(x)$. Hence, the local
concentration profile is $\rho_m(x)= \rho_m g_{m}(x)$.

According to the integral equations method outlined in
section~\ref{theory:a}, the membrane is considered as a the fluid
species labelled as species $\gamma$ (see Fig.~\ref{setup}).
The interaction potential between the membrane and a $j$-species
particle depends only on the particle position, $x$, referred to a
coordinates system set in the middle of the membrane and measured
perpendicularly. Thus, we write $u_{j}({\bf r}_{1})=u_{j}(x)$
which is split as $u_{j}(x)=u_{j}^{el}(x)+u_{j}^{*}(x)$,
being $u_{j}^{el}(x)$ the direct electrostatic potential and
$u_{j}^*(x)$ the hard-core interaction. The former can be found
from Gauss' law, resulting
\begin{equation}
  - \beta u^{el}_{j}(x)= \left\{
    \begin{array}{ll}
    \frac{2 \pi}{\varepsilon}z_{j} e \beta\sigma_{T}(x-L) & \mbox{for
    ${x \ge \frac{d}{2}}$}, \\[2ex]
    \frac{2 \pi}{\varepsilon}z_{j} e \beta \left [\sigma_{T}(-x-L)-
    (\sigma_{1}-\sigma_{2} )d \right ]
    &\mbox{for ${x \le \frac{-d}{2}}$},
    \end{array}
    \right.
\label{electric}
\end{equation}
%%%
%%%
where $L$ is the location of a reference point.
The hard-core interaction is given by
\begin{equation}
  u_{j}^*(x) \; = \; \left\{
    \begin{array}{c@{\quad}l}
      \infty   & {\rm for} \quad \left |x \right |  <  {\displaystyle \frac{d+a}{2}}, \\[2ex]
             0 & {\rm for} \quad \left |x \right | \ge {\displaystyle
             \frac{d+a}{2}},
    \end{array}
   \right.
  \label{hard_term}
\end{equation}
for $j=1,2$. For the impermeable species ($j=3$)
\begin{equation}
  u_{3}^*(x) \; = \; \left\{
    \begin{array}{l@{\quad}l}
      \infty   & \mbox{for $x < {\displaystyle \frac{d+a_3}{2}}$,}\\[2ex]
             0 & \mbox{for $ x \ge {\displaystyle
             \frac{d+a_3}{2}}$}.
    \end{array}
   \right.
  \label{hard_term2}
\end{equation}
This potential imposes $g_{3}(x)=0 \quad {\rm for} \quad x\le
{\displaystyle \frac{d+a_{3}}{2}}$.

In Eq.~(\ref{eq:hnc}) we use the expression of $c_{m j}(r_{13})$
for a primitive model {\em bulk} electrolyte, which has an
analytical expression, written as
%%%
%
\begin{equation}
  c_{m j}(r_{13}) \; =\left\{
  \begin{array}{ll}
  {\displaystyle -\beta u^{el}_{{m j}}(r_{13})=-\beta\frac{ z_m z_j e^2}{\varepsilon r}}&
  \mbox{for $r_{13} \ge a_{mj}$}, \\
  c^{sr}_{m j}(r_{13})+ c^{hs}_{m j}(r_{13})&\mbox{for $r_{13} <
  a_{mj}$},
  \end{array}\right.
  \label{eq:msa}
\end{equation}
%%
%%%%
where $r_{13} \equiv|{\bf r}_{13}|$ is the relative distance
between two ions of species $m$ and $j$. {\fj The particles short
range correlations are considered through the $c^{sr}_{m
j}(r_{13})$ and $c^{hs}_{m j}(r_{13})$. The explicit form of these
functions is given in appendix A. The integral in
Eq.~(\ref{eq:hnc}) can be expressed in cylindrical coordinates,
and analytically calculated in the $\phi$ and $r$ variables (see
Fig.~\ref{setup}), i. e., we consider a cylindrical coordinates
system where $r_{13}^2=x^2+r^2+y^2-2xy$ and $dv_3= d\phi rdr dy $.
After a lengthy algebra, from Eq.~(\ref{eq:hnc}) we get
\cite{lozada84,lozada90b}}
\begin{eqnarray}
g_{j}(x)&=& \exp \left\{\frac{2 \pi}{\varepsilon} z_j e\beta
\left( \sigma_1 + \sigma_2\right)|x| -2\pi A_{j}(x) \right.
\nonumber \\
&+& 2\pi \sum_{m=1}^{2} \rho_m \int_{-\infty}^{-\frac{d+a_m}{2}}
h_{m}(y)G_{mj}(x,y)dy
\nonumber \\
&+&2\pi \sum_{m=1}^{3} \rho_m \int_{\frac{d+a_m}{2}}^{\infty}
h_{m}(y)G_{mj}(x,y)dy  \label{eq:hnc2} \\
&+&2\pi z_j \frac{e^2\beta}{\varepsilon} \sum_{m=1}^{2}z_m\rho_m
\int_{-\infty}^{-\frac{d+a_j}{2}}g_{m}(y)[y+|x-y|]dy
\nonumber \\
 &+&\left.2 \pi z_j \frac{ e^2\beta}{\varepsilon} \sum_{m=1}^{3}
z_m\rho_m
\int_{\frac{d+a_m}{2}}^{\infty}h_{m}(y)[y+|x-y|]dy\right\}.\nonumber
\end{eqnarray}
The first and third integrals include $h_j(y)=g_j(y)-1$ for
particles in the $\alpha$-phase whereas, the second and fourth
integrals, for particles in the $\beta$-phase. Notice the
different summation limits due to the different phase composition.
We have defined
\begin{eqnarray}
  G_{mj}(x,y) &=&  L_{mj}(x,y) + K_{mj}(x,y), \\
  L_{mj}(x,y) &=&
  \int^{\infty}_{|x-y|}c^{sr}_{mj}(r_{13})r_{13}dr_{13}
  ={\frac{e^{2}\beta}{\varepsilon}}D_{mj}(x,y),\label{eq:kernels1}\\
  K_{mj}(x,y) &=&
  \int^{\infty}_{|x-y|}c^{hs}_{mj}(r_{13})\,r_{13}dr_{13},
  \label{eq:kernels2}
\end{eqnarray}
and
\begin{eqnarray}
A_{j}(x)&=&\rho_3 \int^{\frac{d+a_3}{2}}_{-\infty}G_{3j}(x,y)dy \nonumber\\
&+&\sum_{m=1}^{2}\rho_j\int_{-\frac{d+a_m}{2}}^{\frac{d+a_m}{2}}G_{mj}(x,y)dy\nonumber\\
&+& z_j z_3 \rho_3\frac{\beta
e^2}{\varepsilon}\int_{\frac{d+a}{2}}^{\frac{d+a_3}{2}}[y+|x-y|]dy \nonumber\\
&+&z_j e \frac{\beta
d}{\varepsilon}\left(\sigma_1-\sigma_2\right)\Theta(x+d/2),
\end{eqnarray}
with $\Theta(x)$ the step function, defined as
\begin{equation}
\Theta(x) = \left\{
    \begin{array}{c@{\quad}l}
      0 & \mbox{for $x <0$},\\
      1 & \mbox{for $x \ge 0$}.
    \end{array}
\right.
\end{equation}
The expressions for the kernels, $K_{mj}(x,y)$ and $D_{mj}(x,y)$,
are given in appendix A.

From the solution of Eq.~(\ref{eq:hnc2}) one obtains the reduced
concentration profile, $\rho_j(x)=\rho_j g_{j}(x)$. The {\em bulk}
concentrations of species $j$, at the $\alpha$ and $\beta$ phases
($\rho_{j}^{\alpha}$ and $\rho_{j}^{\beta}$) are given by
\begin{equation}
\rho_{j}^{\beta}= \lim_{x \to \infty} \rho_j g_{j}(x),
\label{rho_beta}
\end{equation}
%%%
and
%%%
\begin{equation}
\rho_{j}^{\alpha}= \lim_{x \to -\infty} \rho_j g_{j}(x),
\label{rho_alpha}
\end{equation}
respectively. At the $\beta$ phase $g_{j}(x)\to 1$ as $x\to
\infty$ then $\rho_{j}^{\beta}=\rho_j$. On the other hand, in the
$\alpha$ phase (for $x<0$), ${\displaystyle \lim_{x \to -\infty}
g_{j}(x) \neq 1}$. It must be pointed out that the electrolyte
bulk concentration at the $\alpha$-phase ($\rho_j^\alpha$, for
$j=1,2$) satisfy the bulk electroneutrality condition,
Eq.~(\ref{neutrality}), and are a result from the theory.

\subsection{Computation of the osmotic pressure: contact theorem}
\label{sec:press}

Let us consider a slice of fluid of width $dx$, area of its faces,
A, parallel to the membrane and located at $x$. The force on the
slice in the $x$ direction, $dF_x$, is given by
\cite{olivalesJPC_1980}
\begin{equation}
d{F}_x(x)={E}_x(x)dQ +{S}dp(x) \label{press1}
\end{equation}
being ${E}_x(x)$ the electric field in the $x$ direction at $x$,
$dQ$ the total charge in the fluid slice and $d{p}(x)$ the
pressure difference at the two faces.
Taking into account that
\begin{equation} E_x(x)={-\frac{\partial
\psi(x)}{\partial x}},
\end{equation}
and using Poisson's equation,
\begin{equation}
\frac{\partial^2\psi(x)}{\partial
x^2}=-\frac{4\pi}{\varepsilon}\rho_{el}(x),
\end{equation}
we write
\begin{equation}
dQ=A\rho_{el}(x)dx=-\frac{A\varepsilon}{4\pi}\left(\frac{\partial^2
\psi(x)}{\partial x^2}\right)dx.\end{equation}
Thus, we rewrite Eq.~(\ref{press1}) as
\begin{equation}
d{F}_x(x)=\frac{A\varepsilon}{4\pi} \left(\frac{\partial
\psi(x)}{\partial x}\right)\left(\frac{\partial^2
\psi(x)}{\partial x^2}\right)dx +{A}dp(x),
 \label{balance1}
\end{equation}
or equivalently,
\begin{equation}
d{F}_x(x)=\frac{A\varepsilon}{8\pi}\frac{\partial}{\partial x}
 \left(\frac{\partial \psi(x)}{\partial x}\right)^2dx +{A}dp(x).
 \label{balance}
\end{equation}
Considering that in equilibrium $dF_x(x)=0$ and integrating
Eq.~(\ref{balance}) in the interval $[d/2,\infty)$ with the
boundary condition
\begin{equation}
\lim_{x\to \infty}\frac{\partial \psi(x)}{\partial x}=0,
\end{equation}
we obtain
\begin{equation}
\frac{\varepsilon}{8\pi}
  \left(\frac{\partial \psi(x)}{\partial x}\right)^2_{x=x_0}
 +(p_0^\beta -\Pi^{\beta})=0,
 \label{eq:osm2}
\end{equation}
where $\Pi^{\beta} \equiv \lim_{x\to\infty} p(x)$  is the bulk
fluid pressure and the expression for the pressure on the membrane
right surface, $p_0^\beta\equiv p(0)$, is given by
%%%
%%
\begin{equation}
p_0^\beta  =k_BT \rho_{T}^{\beta}(0)=
k_BT\sum_{i=1}^{3}\rho_ig_{i}\left(\frac{d+a_i}{2}\right),\label{p0}
\end{equation}
where $\rho_{T}^{\beta}(0)=\sum_{i=1}^3\rho_i(\frac{d+a_i}{2})$.
Eq.~(\ref{p0}) is an {\em exact} relationship which can be
obtained by considering the force on the fluid (at the contact
plane) exerted by the {\em hard} wall \cite{chanJCP_1981}. From
basic electrostatics we have
\begin{equation}
\frac{\varepsilon}{4\pi}\left(\frac{\partial \psi(x)}{\partial
x}\right)_{x=d/2} =\int_{d/2}^{\infty}\rho_{el}(x)dx,
\label{eq:maxwell}
\end{equation}
thus, using Eqs.~(\ref{eq:osm2}), (\ref{p0}) and
(\ref{eq:maxwell}) we can write
\begin{equation}
\Pi^{\beta}=\frac{2\pi}{\varepsilon}\left[\int_{d/2}^\infty\rho_{el}(x)dx\right]^2
+k_BT\sum_{i=1}^{3}\rho_ig_{i}\left(\frac{d+a_i}{2}\right),
\label{osmotic1}
\end{equation}
where the first term can be identified as the Maxwell stress
tensor. A similar expression is obtained for the bulk pressure in
the $\alpha$ phase
\begin{equation}
\Pi^{\alpha}=\frac{2\pi}{\varepsilon}\left[\int_{-\infty}^{-d/2}\rho_{el}(x)dx\right]^2
+k_BT\sum_{i=1}^{2}\rho_ig_{i}\left(-\frac{d+a_i}{2}\right).
\label{osmotic2}
\end{equation}
The osmotic pressure, $\Pi$, is defined as
\begin{equation}
\Pi=\Pi^{\beta}-\Pi^{\alpha}. \label{eq:osm}
\end{equation}
From of Eqs.~(\ref{sigma1}), (\ref{sigma2}), (\ref{osmotic1}) and
(\ref{osmotic2}), Eq.~(\ref{eq:osm}) becomes,
\begin{eqnarray}
\Pi&=&\frac{2\pi}{\varepsilon}\left\{[\sigma^\beta]^2-
[\sigma^\alpha]^2\right\} +k_BT\sum_{i=1}^{3}\rho_ig_{i}\left(\frac{d+a_i}{2}\right)\nonumber \\
&-&k_BT\sum_{i=1}^{2}\rho_ig_{i} \left(-\frac{d+a_i}{2}\right).
\label{contact:eq}
\end{eqnarray}
%%%
%%
%%
Due to the fluid-fluid correlation across a thin membrane, the
induced charge densities($\sigma^\alpha$ and $\sigma^\beta$) and
$g_{i}\left(\pm\frac{d+a_i}{2}\right)$ depend on the fluid
conditions ($\rho_i$, $z_i$, $a_i$, with $i=1,... ,3$) and on the
membrane parameters ($\sigma_1$, $\sigma_2$ and $d$)
\cite{lozadaPRL_1996,lozadaPRE_1997,aguilar2001}. The computed
value of $\Pi$ (using $\sigma^\alpha$, $\sigma^\beta$ and
$g_{i}\left(\pm\frac{d+a_i}{2}\right)$ from HNC/MS), however, does
{\em not} depend on the membrane parameters. We have numerically
corroborated this fact by computing $\Pi$ for several values of
$\sigma_1$, $\sigma_2$ and $d$. This is physically appealing since
the pressure can only depend on the bulk fluid conditions at both
sides of the membrane. However, we had to do very precise
calculations of $g_{i}(x)$, particularly in the neighborhood of
$x=\pm \frac{d+a_i}{2}$, to prove the above statement.

Eq.~(\ref{contact:eq}) is an {\em exact} theorem to compute the
osmotic pressure, $\Pi$, in terms of microscopic quantities. A
similar expression for the osmotic pressure was derived by Zhou
and Stell \cite{stell88_2}. However, the differences between the
current derivation and that of those authors are due to the
ions-membrane short range interactions. If we use, in the Zhou and
Stell theory, the hard wall interaction between the permeable ions
and the membrane (provided by Eq.~(\ref{hard_term})) we recover
Eq.~(\ref{contact:eq}).
%%

%\newpage
%Fig2
%%%%%%%%%%%%%%%%%%%%%%%%%%%%%%%%%%%%
\begin{figure}
\includegraphics[width=8cm]{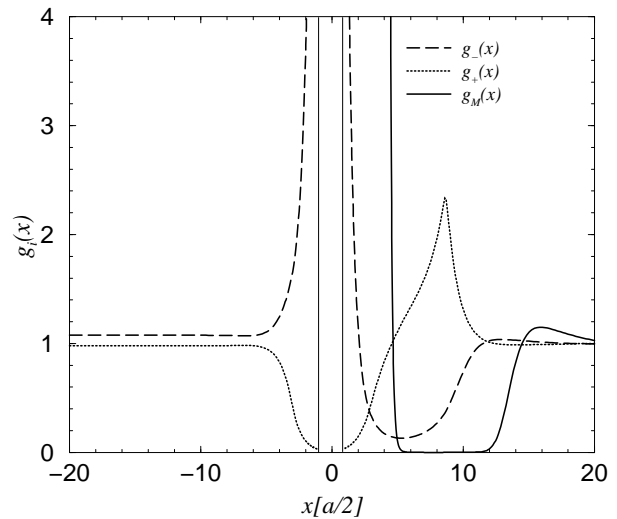}
\caption{Reduced concentration profiles (RCPs) for a macroions
solution ($\rho_M=0.01$M, $z_M=-10$) in a monovalent electrolyte
($\rho_+=1.1$M and $\rho_-=1.0$M), with $a_3=3.8a$,
$\sigma_1=\sigma_2=0.272$ \Cmm and $d=a$. The continuous, dashed
and dotted lines represent the RCPs for the macroions, anions and
cations, respectively.} \label{monovalent_pstveM_3.8}
\end{figure}
%%%%%%%%%%%%%%%%%%%%%%%%%%%%%%%%%%%%

\section{Results and discussion}
\label{results}

Several physical effects determine particles adsorption on the
charged membrane. One of the most relevant is the
membrane-particle direct interaction energy, which, at the surface
is given by

\begin{equation}
U_i=q_i u_{i}\left(\frac{a_i}{2}\right)
   =\frac{2\pi
   q_i\sigma}{\varepsilon}\left(L-\frac{a_i}{2}\right),
\label{surface}
\end{equation}
being $L$ the location of a reference point. The {\fj more
negative} the value of $U_i<0$, particles adsorption is
energetically more favorable. Many body correlations play also an
important role in the adsorption phenomena and are responsible for
the surface-particle forces of non-electrostatic origin.
Although it is not possible to sharply distinguish the origin of
correlations, we assume that the particles volume fraction
($\eta_T\equiv\frac{\pi}{6}\sum_{i}\rho_i a_{i}^3$) {\fj
quantifies the contribution of short range correlations. To
quantify the effect of the coulombic interaction (long range
correlations), we define the parameter $\xi_{ij}=\beta
q_{i}q_{j}/\varepsilon a_{ij}$ and we simply write
$\xi_{i}\equiv\xi_{ii}=\beta q_{i}^2/\varepsilon a_{i}$, when
$i=j$. In the discussion, the effects produced by varying the
fluid parameters  ($a_i$, $z_i$ and $\rho_i$, with $i=1,...,3$)
will be associated with the increment (or decrement) of the
contributions arising from short and long range correlations
($\eta_T$ and $\xi_{ij}$).}

%Many body may induce {\em charge reversal} (CR) of a charged
%surface in contact with a multivalent ions solution.

{\fj For our discussion it is useful remark the following
concepts: When the amount of adsorbed charge exceeds what is
required to screen the surface charge, it is said to occur a
surface {\em charge reversal} (CR). In consequence,} at certain
distance from the surface the electric field is inverted. Next to
the CR layer, a second layer of ions (with same charge sign as
that of the surface) is formed, producing a {\em charge inversion}
(CI) of the electrical double layer
\cite{attard_1995,kjellander_1998}. Such a denomination (charge
inversion) is originated from the fact that ions invert their role
in the diffuse layer. In the past, we have shown that charge
reversal and charge inversion are many body effects and are
induced by a compromise between short range correlations
($\eta_T$) and electrostatic long range correlations ($\xi_{ij}$)
\cite{jimenezPRL_2003,jimenezJCP_2003}.

%%%%%%%%%%%%%%%%%%%%%%%%%
%\newpage
\begin{figure}
\includegraphics[width=8cm]{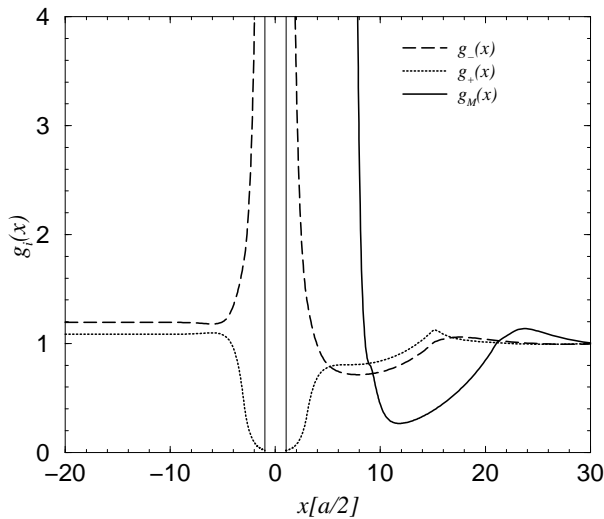}
\caption{Same as in Fig.~\ref{monovalent_pstveM_3.8} but with
$a_3=7a$. The lines meaning is the same as in
Fig.~\ref{monovalent_pstveM_3.8}} \label{monovalent_pstveM_7.0}
\end{figure}
%%%%%%%%%%%%%%%%%%%%%%%%%%
%\newpage

The HNC/MS equations for a semipermeable membrane,
Eqs.~(\ref{eq:hnc2}), are numerically solved using a finite
element technic \cite{lozada90a,mier}. From the solution of HNC/MS
equations the reduced concentration profiles (RCPs), $g_{j}(x)$,
are obtained. In the discussion, we adopted the following
notation: $z_1=z_+$, $z_2=z_-$ and $z_3=z_M$ for the number
valence of cations, anions and macroions, respectively; idem  for
$\rho_i$ and $a_i$. In all our calculations we have used fixed
values of $T=298$K, $\varepsilon=78.5$ and $a=4.25$~{\AA}. The
effects of salt valence, macroions size, and membrane surface
charge density on macroions adsorption are analyzed.
Thus, we consider macroions ($\rho_M=0.01$M and $z_M=10$) in a (a)
monovalent ($\rho_+=1.1$M and $\rho_-=1.0$M) and (b) divalent
($\rho_+=0.55$M and $\rho_-=0.5$M) electrolyte solutions.
%These parameters are summarized in Tab.~\ref{tab:system_c}.
For each case, we considered two macroion sizes ($a_M=3.8a$ and
$a_M=7a$) and three values for the membrane charge densities: i)
$\sigma_1=\sigma_2= 0.272${\Cmm}, ii) $\sigma_1=\sigma_2=
-0.272${\Cmm} and iii) $\sigma_1=0.68$\Cmm, $\sigma_2=
-0.136${\Cmm}. Finally we show calculations for the osmotic
pressure (as a function of the macroion concentration) compared
with experimental results.

%Fig4
%%%%%%%%%%%%%%%%%%%%%%%%%%%%%%%
\begin{figure}
\includegraphics[width=8cm]{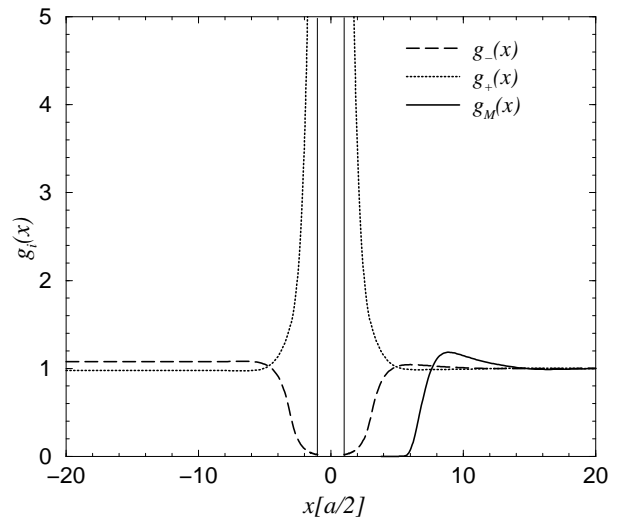}
\caption{Reduced concentration profiles (RCPs) for a macroions
solution ($\rho_M=0.01$M, $z_M=-10$) in a monovalent electrolyte
($\rho_+=1.1$M and $\rho_-=1.0$M), with $a_3=3.8a$,
$\sigma_1=\sigma_2=-0.272$ \Cmm and $d=a$. The lines meaning is
the same as in Fig.~\ref{monovalent_pstveM_3.8}}
\label{monovalent_ngtveM_3.8}
\end{figure}
%%%%%%%%%%%%%%%%%%%%%%%%%%%%%%
%\newpage
%%

\subsection{Macroions in a monovalent electrolyte ($z=1$)}
In this subsection we discuss the case of macroions ($z_M=-10$) in
a monovalent electrolyte solution ($\rho_+=1.1$M, $\rho_-=1.0$M,
$z_+=-z=1$).
\subsubsection{Positively charged membrane}

In Fig.~\ref{monovalent_pstveM_3.8} we present the RCPs when the
membrane is positive and symmetrically charged
($\sigma_1=\sigma_2=0.272$\Cmm), the membrane thickness is $d=a$
and $a_M=3.8a$. As a result, we obtained the asymptotic values of
the distribution functions in the $\alpha$-phase,
$g_{-}(-\infty)=1.0758$ and $g_{+}(-\infty)=0.9779$, such that
Eq.~(\ref{neutrality}) is satisfied. In both phases we observe
that the negative ions are adsorbed and the positive ions
expelled, as it is expected. In the $\beta$-phase we observe that
the adsorption of macroions is more favorable than the adsorption
of negative small ions. This is understood in terms of the
electrostatic energy of  one particle of species $i$ at the
membrane surface, $U_i$: since $|z_M|>|z_-|$, from
Eq.~(\ref{surface}) it is easy to see that $U_{M}<U_-<0$, which
favors macroions adsorption. As we pointed out above, many body
correlations also influence adsorption. In this case
macroion-macroion long range correlations are predominantly more
important than the ion-ion correlations  since $\xi_{M} \approx 7
\xi_{-}$. {\fj Short range correlations are also important and
play an important role in the macroions adsorption: an increment
of macroions concentration (keeping constant $a_M$ and $z_M$)
implies an increment of $\eta_T$ and produces an increment of the
macroions adsorption. However an increment of $\eta_T$ not always
is followed by an increase of the adsorption as it will be noticed
in the discussion of Fig.~\ref{monovalent_pstveM_7.0}}. In the
macroions RCP we find a first macroions layer next to the membrane
surface, after this layer there is a region where the macroions
are completely expelled and then a small second peak. The negative
small ions are also adsorbed to the membrane's surface but their
concentration is smaller than for the macroions. The RCP for
positive ions show that these are completely expelled from the
membrane right surface. However, a peak in the RCP is found at
$x\approx 4.3 a$ which is the ion-surface distance when there is a
macroion in between. This peak implies an effective surface-cation
attraction due to a {\em field inversion} caused by the surface
CR.

Fig.~\ref{monovalent_pstveM_7.0} shows the results obtained for
the same system as in Fig.~\ref{monovalent_pstveM_3.8} except that
the macroions are larger, $a_M=7a$. At the $\alpha$-phase the
distribution function have the same qualitative behavior as in
Fig.~\ref{monovalent_pstveM_3.8}, however, in this case
$g_{-}(-\infty)=1.1938$ and $g_{+}(-\infty)=1.0853$, i. e., the
amount of salt in the $\alpha$-phase increases by increasing the
macroions size. Respect to Fig.~\ref{monovalent_pstveM_3.8}, a
decrement of macroions adsorption is observed in
Fig.~\ref{monovalent_pstveM_7.0}. The increment of the macroions
diameter implies a decrement of the macroions-macroions long range
correlations (in this case $\xi_{M}\approx 2\xi_{-}$) and an
increment of $\eta_T$, {\fj and hence, an increment of the
contributions arising from correlations of short range nature.
Such an increment, however, does {\em not} increase (but decrease)
the macroions adsorption}. This is understood in terms of
Eq.~(\ref{surface}) [$U_M(a_M=3.8a) < U_M(a_M=7a)$] which implies
that the adsorption is energetically less favorable for $a_M=7$
than for smaller macroions with the same charge.

\begin{figure}
\includegraphics[width=8cm]{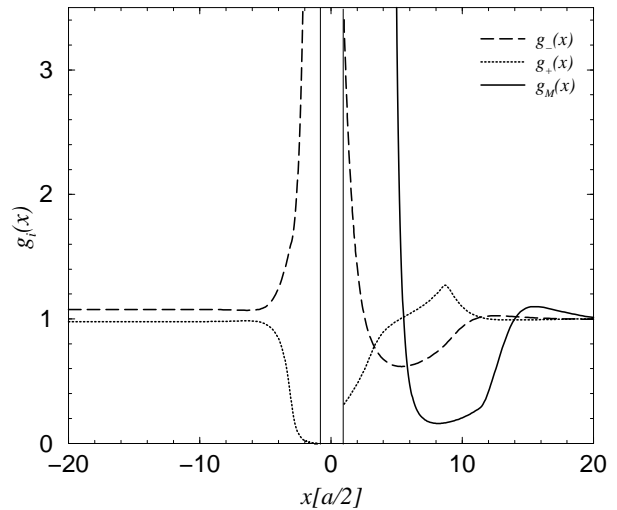}
\caption{Reduced concentration profiles (RCPs) for a macroions
solution ($\rho_M=0.01$M, $z_M=-10$) in a monovalent electrolyte
($\rho_+=1.1$M and $\rho_-=1.0$M), with $a_3=3.8a$,
$\sigma_1=0.68$ \Cmm, $\sigma_2=-0.136$ \Cmm and $d=a$. The lines
meaning is the same as in Fig.~\ref{monovalent_pstveM_3.8}.}
\label{correlation11}
\end{figure}
%%%%%%%%%%%%%%%%%%%%%%%%

\subsubsection{Negatively charged membrane}

Fig.~\ref{monovalent_ngtveM_3.8} shows the RCPs when the membrane
is negative and symmetrically charged
($\sigma_1=\sigma_2=-0.272$\Cmm) for $a_M=3.8a$.  The membrane
thickness is $d=a$. The asymptotic value of the distributions
function does not depend on the membrane's charge and thickness,
hence, the $g_{i}(-\infty)$ have the same value as in
Fig.~\ref{monovalent_pstveM_3.8}. The RCP for small ions behave in
a normal way in the sense that positive ions are attracted to the
to the membrane surface, whereas the negative ions are expelled
from it. In the $\beta$-phase, it is observed that macroions are
expelled from the membrane surface for $x< 6(a/2)$. After this
zero concentration region, an small peak in the macroions RCP is
observed at $x\approx 9(a/2)$ indicating an effective
macroion-membrane {\em attractive} force and an slight surface CR
produced by the small cations. At these electrolyte conditions
($1:1$ and $\rho_+=1$M with no macroions), cations display a
monotonically decaying distribution profile \cite{lozada82},
hence, the oscillatory behavior of the RCPs for the small ions (in
the $\beta$-phase) is a consequence of the presence of macroions.
By considering the macroions size $a_M=7.0$ (not shown), the
qualitative behavior of the RCPs is similar to that of
Fig.~\ref{monovalent_ngtveM_3.8} with the following differences:
(i) the RCPs oscillations of the small ions at the $\beta$-phase
are of longer range and (ii) the macroions RCP maximum is higher.
{\fj This fact points out the relevance of the effect of the
particles size and concentration (particles volume fraction) in
the effective attraction between like charged particles in
solution
\cite{muthuJCP_1996,crockerPRL_1999,likecharges1,likecharges2,likecharges4}.}

%\newpage
%%%%%%%%%%%%%%%%%%%%%%%%%%%%%%%%%
\begin{figure}
\includegraphics[width=8cm]{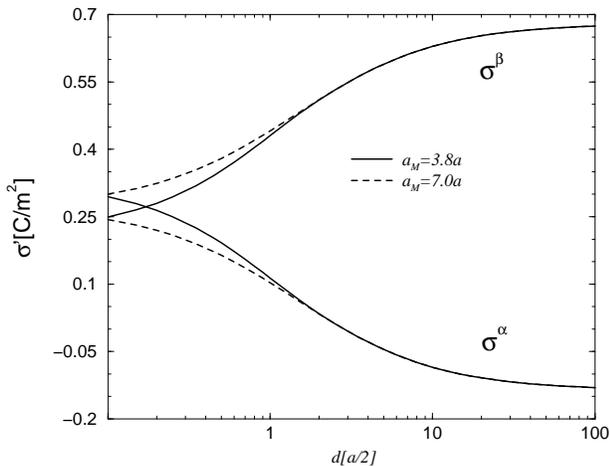}
\caption{Induced charge densities as a function of the membrane
thickness for a macroions solution ($\rho_M=0.01$M, $z_M=-10$) in
a monovalent electrolyte ($\rho_+=1.1$M and $\rho_-=1.0$M) with
$\sigma_1=0.68$ \Cmm and $\sigma_2=-0.136$ \Cmm. The continuous
line represents calculations for $a_{M}=3.8a$, whereas the dashed
line for $a_{M}=7.0a$} \label{induced_charges-1}
\end{figure}
%%%%%%%%%%%%%%%%%%%%%%%%%%%%%%%%%%
%\newpage
%%%%%%%%%%%%%%%%%%%%

From the analysis presented from Fig.~\ref{monovalent_pstveM_3.8}
to \ref{monovalent_ngtveM_3.8} we see that macroions adsorption
increases by increasing $\eta_T$, $\xi_M$ and $-U_M$. However, we
point out the following findings: when macroions and the surface
are oppositely charged, long range electrostatic correlations
dominate over short range correlations. Hence, adsorption is
enhanced by increasing $\xi_{M}$ and/or $-U_M$. On the other hand,
when macroions and the surface are like charged the mechanism for
macroions adsorption (mediated by the small ions) is mainly driven
by short range correlations, thus, adsorption increases by
increasing $\eta_T$ even though $\xi_M$ decreases.

\subsubsection{Unsymmetrically charged membrane}

In Fig.~\ref{correlation11} it is shown the RCPs at the two
membrane sides for a macroions diameter $a_{3}=3.8a$ and $d=a$.
The membrane is {\em unsymmetrically} charged with $\sigma_1=0.68$
\Cmm and $\sigma_2=-0.136$ \Cmm. At the right hand side surface
macroions and small anions (negatively charged) are adsorbed on
the membrane, in spite of $\sigma_2<0$. The adsorption of
negatively charged particles on a negatively charged surface is
due to the correlation between the two fluids. This is understood
by considering the following two facts: (i) the membrane has a
positive {\em net} charge ($\sigma_T>0$) and (ii) the charge on
the left hand side surface ($\sigma_1$) is not completely screened
by the excess of charge in its corresponding fluid phase
($\sigma^{\alpha}$), therefore, the electric field produced by
$\sigma_1+\sigma^{\alpha}$ overcomes the field produced by
$\sigma_2$, inducing macroions and anions adsorption. From here,
we observed that $\sigma^\alpha$ and $\sigma^\beta$ depend on $d$
as it is discussed below.

In Fig.~\ref{induced_charges-1} it is shown the excess of charge
densities $\sigma_\alpha$ and $\sigma_\beta$, as a function of the
membrane's thickness $d$. The dependence on the wall thickness of
the induced charge densities $\sigma_{\alpha}$ and
$\sigma_{\beta}$, is a manifestation of the correlation between
the fluids. The correlation between the two fluids is due to the
electrostatic interaction among the particles at both phases but,
more importantly, to the fact that they are at constant chemical
potential. For a sufficiently large membrane's thickness the
induced charge density in each fluid phase screens its
corresponding membrane surface, i. e., $\sigma^\alpha \to
-\sigma_1$ and $\sigma^\beta \to -\sigma_2$ as $d \to \infty$. At
$d=100a$ each fluid has screened its respective surface charge
density. Here we show a comparison between the results obtained
for $a_{M}=3.8a$ and $a_{M}=7a$. In both cases the results are
qualitatively similar.

%Fig7
%%%%%%%%%%%%%%%%%%%%%
\begin{figure}
\includegraphics[width=8cm]{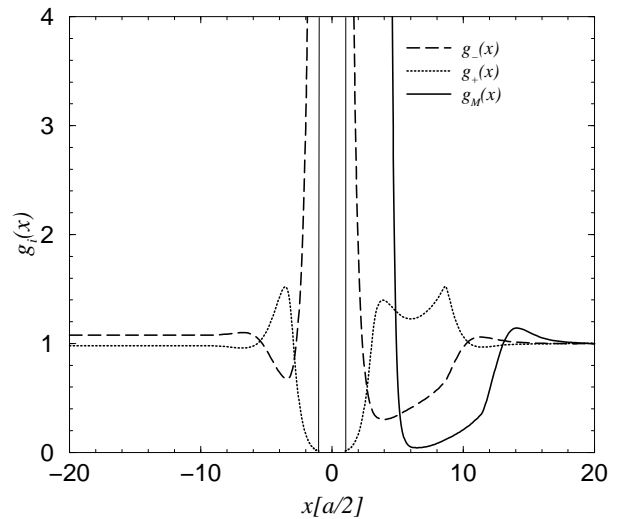}
\caption{Reduced concentration profiles (RCPs) for a macroions
solution ($\rho_M=0.01$M, $z_M=-10$) in a divalent electrolyte
($\rho_+=0.55$M and $\rho_-=0.5$M), with $a_3=3.8a$,
$\sigma_1=\sigma_2=0.272$ \Cmm and $d=a$. The continuous, dashed
and dotted lines represent the RCPs for the macroions, anions and
cations, respectively.} \label{divalent_pstveM_3.8}
\end{figure}
%%%%%%%%%%%%%%%%%%%%%%
%\newpage

\subsection{Macroions in a divalent electrolyte ($z=2$)}
We now discuss the case of macroions ($z_M=-10$) in a divalent
electrolyte solution ($\rho_+=0.55$M, $\rho_-=0.5$M, $z_+=-z=2$).

\subsubsection{Positively charged membrane}

In Fig.~\ref{divalent_pstveM_3.8} we show the RCPs for
$\sigma_1=\sigma_2=0.272$\Cmm, $d=a$ and $a_M=3.8a$. At the
$\alpha$-phase we observe oscillations of the RCPs, which is a
typical behavior of a divalent electrolyte. Although we observe a
strong adsorption of macroions in the $\beta$-phase, the amount of
adsorbed small negative ions is even larger (the concentrations of
macroions and small negative ions at the interface are
$\rho_{M}\left(\frac{d+a_M}{2}\right)\approx 7.2$M  and
$\rho_{-}\left(\frac{d+a}{2}\right)\approx 16$M, respectively).
Energetically, macroions adsorption should be more favorable,
however, macroions adsorption is inhibited because divalent
positive ions more efficiently screen macroion-membrane and
macroion-macroion interactions. We infer this from the RCP for
cations, which displays two small peaks at $x \approx 2a$ and
$x\approx 4.3a$. {\fj The first peak corresponds to a positive
ions layer contiguous to the negative ions adsorbed on the wall.
The position of the second peak corresponds to a positive ions
layer next to the macroions layer. This structure indicates that
positive ions surround macroions due to their strong electrostatic
interaction, in this case $\xi_{M+}\approx -8.3
\frac{e^2}{\varepsilon k_B T a}\approx -14$, whereas for macroions
in a monovalent electrolyte $\xi_{M+}\approx -7$. By comparison of
Figs.~\ref{monovalent_pstveM_3.8} and \ref{divalent_pstveM_3.8} we
see that} macroions (next to an oppositely charged surface) are
better adsorbed when they are in a monovalent solution rather than
in a multivalent solution: for this particular case of macroions
($\rho_M=0.01$M, $z_M=-10$ and $a_M=3.8a$),
$\rho_M(\frac{d+a_M}{2})= 20$M when macroions are in a monovalent
electrolyte, whereas $\rho_{M}\left(\frac{d+a_M}{2}\right)\approx
7.2$M when macroions are in a divalent electrolyte.

%Fig8
%%%%%%%%%%%%%%%%%%%%%%%%%%
\begin{figure}
\includegraphics[width=8cm]{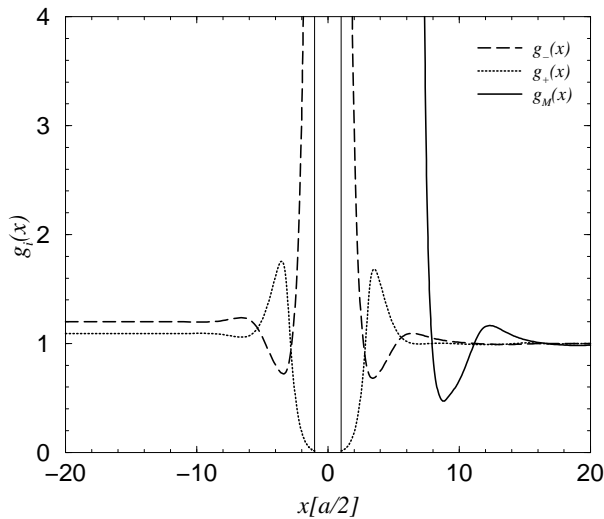}
\caption{Same as in Fig.~\ref{divalent_pstveM_3.8} but with
$a_3=7a$. The lines meaning is the same as in
Fig.~\ref{divalent_pstveM_3.8}.} \label{divalent_pstveM_7.0}
\end{figure}
%%%%%%%%%%%%%%%%%%%%%%%%%%
%\newpage
%Fig9
%%%%%%%%%%%%%%%%%%%%%%%%%

In Fig.~\ref{divalent_pstveM_7.0} we show the RCPs for the same
conditions as in Fig.~\ref{divalent_pstveM_3.8} but $a_M=7a$.
Although macroions adsorb in the $\beta$-phase, they do not {\fj
influence} significantly on the local concentration of small ions.
Hence, we see that the RCPs for small ions are quantitatively
similar in both phases: the concentrations of counterions at the
membrane surfaces are $\rho\left(\pm\frac{d+a}{2}\right)\approx
22.5$M, in addition, the RCPs maxima are located symmetrically
around $x\approx \pm 1.7a$. The adsorption of macroions decrease
(respect to Fig.~\ref{divalent_pstveM_3.8}) due to the efficient
screening of the membrane charge by the small negative ions and
because the adsorption of larger ions (keeping $z_M$ constant) is
energetically less favorable, as it was pointed out in the
discussion of Fig.~\ref{monovalent_pstveM_7.0}. {\fj We see that
the layer of cations around macroions (seen in
Fig.~\ref{divalent_pstveM_3.8}) desapears due to the decrement of
their coulombic interaction, in this case $\xi_{M+} \approx
-7.6$.}

\begin{figure}
\includegraphics[width=8cm]{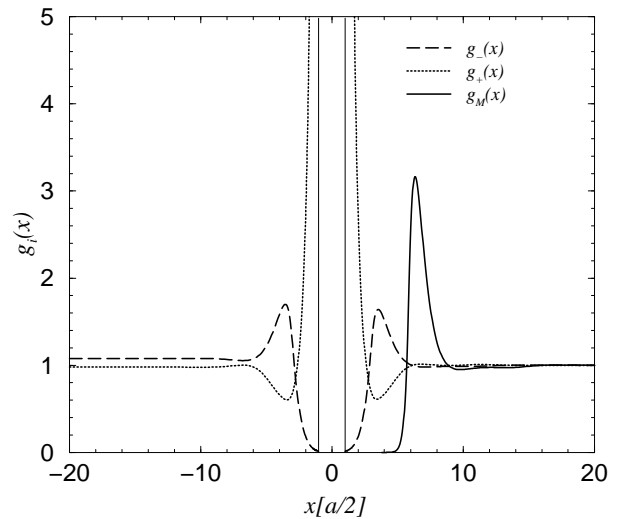}
\caption{Reduced concentration profiles (RCPs) for a macroions
solution ($\rho_M=0.01$M, $z_M=-10$) in a divalent electrolyte
($\rho_+=0.55$M and $\rho_-=0.5$M), with $a_3=3.8a$,
$\sigma_1=\sigma_2=-0.272$ \Cmm and $d=a$. The lines meaning is
the same as in Fig.~\ref{divalent_pstveM_3.8}.}
\label{divalent_ngtveM_3.8}
\end{figure}
%%%%%%%%%%%%%%%%%%%%%%%%%
%\newpage
%%

\subsubsection{Negatively charged membrane}

In Fig.~\ref{divalent_ngtveM_3.8} we show the RCPs for a
negatively charged membrane ($\sigma_1=\sigma_2=-0.272$\Cmm) with
$d=a$ and $a_{M}=3.8a$. The quantitative behavior of the small
ions RCPs is similar at both phases, as much as the location of
the maxima located symmetrically at $x \approx \pm 1.8a$. At the
$\beta$-phase, the macroions RCP displays a peak at $x=3.2a$
indicating the formation of a macroions layer. The {\em
adsorption} of macroions at this layer is enhanced (respect to
Fig.~\ref{monovalent_ngtveM_3.8}) because of the surface {\em
charge reversal} produced by the divalent cations, i. e.,
macroions see the membrane surface with an effective positive
charge. By increasing macroions size (keeping $z_M$ constant),
macroions attraction to an oppositely charged surface is
energetically less favorable, thus, we observe that macroions
adsorption decreases by increasing macroions size, in opposition
to the behavior of macroions in a monovalent electrolyte.

It seems to be a general feature that macroions next to an
oppositely charged surface are better adsorbed when they are in a
monovalent solution than in a multivalent solution. {\fj It is
important to point out the formation of a cations layer around the
macroions when $|\xi_{M+}|$ is high (seen for divalent electrolyte
in Fig.~\ref{divalent_pstveM_3.8} )}. On the other hand, when
macroions and the surface are like charged, a {\em divalent}
electrolyte solution mediates an effective membrane-macroions
attraction, this attraction is favored by a higher value of
$\xi_{M}$ rather than for a high value of $\eta_M$, as in the
monovalent case.

%Fig10
%%%%%%%%%%%%%%%%%%%%%%
\begin{figure}
\includegraphics[width=8cm]{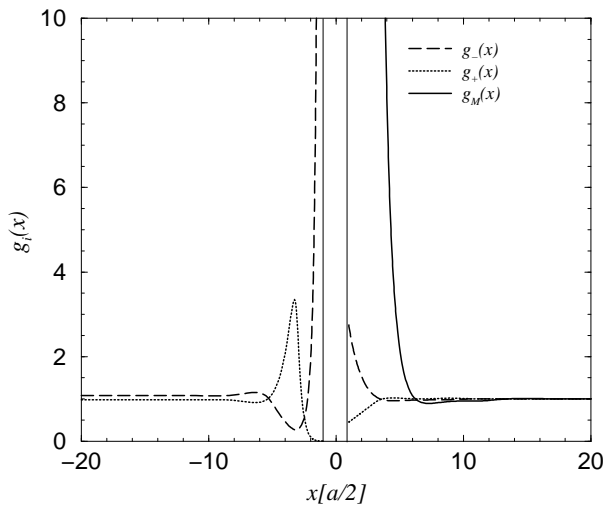}
\caption{Reduced concentration profiles (RCPs) for a macroions
solution ($\rho_M=0.01$M, $z_M=-10$) in a divalent electrolyte
($\rho_+=0.55$M and $\rho_-=0.5$M), with $a_3=3.8a$,
$\sigma_1=0.68$ \Cmm, $\sigma_2=-0.136$ \Cmm and $d=a$. The lines
meaning is the same as in Fig.~\ref{divalent_pstveM_3.8}.}
\label{correlation22}
\end{figure}
%%%%%%%%%%%%%%%%%%%%%%%

\subsubsection{Unsymmetrically charged membrane}

In Fig.~\ref{correlation22} we show the RCPs at the two phase for
$\sigma_1=0.68$\Cmm, $\sigma_2=-0.136$\Cmm, $a_{3}=3.8a$ and
$d=a$. The correlation between the two fluids is manifested by the
attraction of negatively charged particles towards the negatively
charged surface at the {\fj $\alpha$-phase}. However, the
adsorption of macroions is quite less efficient than in the
monovalent electrolyte case of Fig.~\ref{correlation11}: In the
monovalent case the contact value of the local concentration is
$\rho_M(\frac{d+a_M}{2}) \approx 2.7$M whereas in this case it is
$\rho_M(\frac{d+a_M}{2}) \approx 0.12$M. This is due mainly to the
more efficient field screening by the divalent electrolyte at the
left hand side surface. In the monovalent electrolyte case the
induced charge density at the $\alpha$-phase is
$\sigma_{\alpha}=-0.43$\Cmm when $d=a$ whereas in the divalent
electrolyte case $\sigma_{\alpha}=-0.50$\Cmm.

\subsection{The osmotic pressure}

\begin{figure}
\includegraphics[width=8cm]{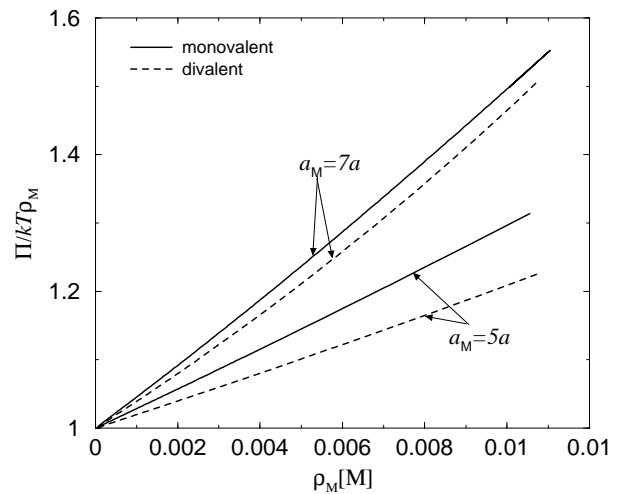}
\caption{Reduced osmotic pressure as a function of the macroions
concentration, $\rho_{\rm M}$, for $z_{\rm M}=-10$ and two
macroion sizes, $a_{\rm M}=5a$ and $a_{\rm M}=7a$. The solid line
represents the osmotic pressure for macroions in a monovalent
electrolyte ($\rho_-=1.0$M) whereas the dashed line for macroions
in a divalent electrolyte ($\rho_-=0.5$M).} \label{size-osmotica}
\end{figure}

Although the adsorption of macroions is strongly influenced by the
membrane surface charge and thickness, we observe however, that
the osmotic pressure does not depends on these membrane properties
as it has been pointed out in subsection~\ref{sec:press}.  In
Fig.~\ref{size-osmotica} we show the osmotic pressure (obtained
from HNC/MS theory) as a function of the macroions concentration
$\rho_{\rm M}$, for two macroion sizes. This plot shows the
osmotic pressure for macroions in a monovalent electrolyte and in
a divalent electrolyte. The osmotic pressure increases by
increasing the particles excluded volume (either $\rho_{\rm M}$ or
$a_{\rm M}$), on the other hand, it is not observed a qualitative
difference between the curves for the osmotic pressure of
macroions in a monovalent and divalent electrolytes.

%%%%%%%%%%%%%%%%%%%%%%%

\begin{figure}
\includegraphics[width=8cm]{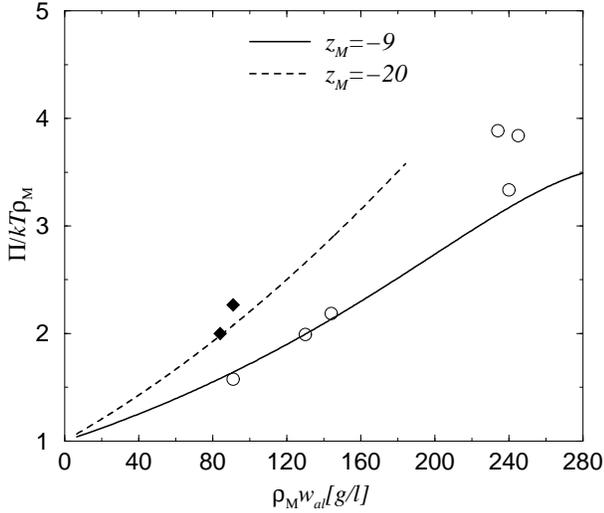}
\caption{Bovine serum albumin reduced osmotic pressure as a
function of albumin weight concentration, $\rho_Mw_{al}$, at
25$^{o}$C and in 0.15M NaCl at pH=7.4($Q=-20e$) and pH=5.4($-9e$)
from ref.~\cite{vincent80}. The curves are obtained from
Eq.~(\ref{osmotica}) using $a_M=62${\AA}, $a=4.25${\AA} and
$z=1$.} \label{osmotica}
\end{figure}

Computation of the osmotic pressure of proteins solutions is an
issue addressed by some authors \cite{youseff98,deserno2002}.
Results of particular interest are those for albumin solutions,
for which theoretical calculations and measurements of the osmotic
pressure have been reported. In Fig.~\ref{osmotica} we show the
osmotic pressure predictions of HNC/MS theory (as a function of
the protein concentration) and experimental results.
The experimental data correspond to a solution of albumin in a
$0.15$M NaCl aqueous solution. In accordance with titration
measurements, the albumin has a net charge of $Q=-9e$ and $Q=-20e$
for pH$\approx 5.4$ and pH$\approx 7.4$, respectively, in our
calculations we have used $a_M=62${\AA} which corresponds to the
experimental protein diameter. It is remarkable the excellent
agreement between theory end experiment as well as the fact that
no adjustable parameters have been used.
The prediction of HNC/MS fits well the experimental data even for
protein concentration as high as $30\%$ the protein volume in
solution (which is estimated assuming the albumin molecular weight
$w_{al}=69$Kg/mol). For higher protein concentration, HNC/MS shows
discrepancies with experimental measurements which may be
associated with the following facts: (i) the albumin molecule is
not {\em spherically} symmetric as in the model, therefore, the
protein geometry becomes relevant when the protein volume fraction
is high, (ii) integral equations are approximated theories,
meaning that they do not take into account all the particle
correlations.
%%

%\subsection{Discussion}

%

%%%%%%%%%%%%%%%%%%%%%%%

\section{Conclusions}
\label{conclusions}

{\fj We studied a model charged membrane separating two fluid
phases ($\alpha$ and $\beta$). The $\beta$-phase phase contains
macroions in an electrolyte solution and the $\alpha$-phase is a
simple electrolyte solution. The system is modeled in such a way
that the small ions at both phases are at the same chemical
potential, thus, the membrane is considered to be semipermeable.
It is important to point out that we cionsidered explicitly the
effect of particles size (short range correlations) and
electrostatic long range correlations. Here we have applied the
hypernetted chain/mean spherical integral equations to the
semipermeable membrane model. Also we have derived the equation
for the osmotic pressure throug a simple forces balance.}

By solving the HNC/MS integral equations we obtained the particles
concentration profiles which allowed us to study the adsorption of
macroions on the membrane. We analyzed the influence of several
factors in the macroions adsorption: membrane surface charge,
membrane thickness, the effect of salt and macroions size. From
this study we emphasize the following results:
1) When the membrane and macroions are oppositely charged the
adsorption is energetically favorable, however, if multivalent
ions are present the membrane-macroions interaction is screened
which is unfavorable for macroions adsorption. On the other hand,
the larger the macroions (keeping the same charge and
concentration) the adsorption is energetically less favorable. 2)
The attraction of macroions towards a like-charged surface seems
to be energetically unfavorable. Nevertheless, in our model, we
find that such an attraction is feasible and it is due to short
range correlations, which are properly considered in our theory.
These previous results are not a consequence of the permeability
condition but are general. 3) As a consequence of the permeability
condition and constant chemical potential, we have that for an
unsymmetrically charged membrane the fluids {\em correlation} may
produce adsorption of charged macroions on a like charged surface.
In addition, the permeability condition implies a non trivial
relation between the induced charge densities ($\sigma^\alpha$ and
$\sigma^\beta$) and the membrane thickness. The theory predictions
are robust as it is shown by the excellent agreement between
theory and experiment where we have not used adjustable
parameters. The results of this work could be technologically
relevant for the design of selective membranes \cite{membranes}.

\section{Acknowledgments}

We gratefully acknowledge the financial support of INDUSTRIAS
NEGROMEX.
\appendix
\section{Appendix}

\subsection{The Mean Spherical Closure}
%%%

The primitive model is the simplest model for an electrolyte that
includes many relevant aspects of real solutions. In the general
case the primitive model is constituted by $n$-species of
particles, with the mixture is embedded in a uniform medium of
dielectric constant $\varepsilon$ at temperature $T$. Each species
is defined by the particles point charge at the center, $q_i=z_i
e$ (where $e$ stands for the proton's charge and $z_i$ for the
ionic valence), the ionic diameter, $a_i$, and number
concentration, $\rho_i$. The fluid is constrained to the following
condition
\begin{equation}
\sum_{i=1}^nz_i\rho_i=0.
\end{equation}

The expressions for the direct correlation functions,
$c_{ij}({r}_{13})$, for a {\em bulk} electrolyte (required in n
Eq.~(\ref{eq:hnc})) were obtained by Blum \cite{blum75} and
Hiroike \cite{hiroike77}, through the MS closure, and are written
as
%%%
\begin{equation}
\label{eq1} c_{ij} (r_{13}) = \frac{e^{2}\beta}{\varepsilon}d_{ij}
(r_{13}) + c_{ij}^{hs} (r_{13}) - \beta {\frac{{z_{i} z_{j}
e^{2}}}{{\varepsilon r_{13}}}},
\end{equation}
with $c_{ij}^{sr}(r_{13})={\displaystyle
\frac{e^{2}\beta}{\varepsilon}}d_{ij}(r_{13})$, $\beta =1/k_BT$
and
%%%
\begin{equation}
\begin{array}{lll}
d_{ij}(r_{13}) = \\
\left\{\begin{array}{l} \displaystyle{b^{(1)}_{ij} +
\frac{z_{i} z_{j}}{r_{13}}}, \;\mbox{for $0 \le r_{13} \le \lambda_{ij}$},\\
\displaystyle{\frac{b^{(2)}_{ij} + z_{i} z_{j}}{r_{13}} -
b^{(3)}_{ij} + b^{(4)}_{ij} r_{13} + b^{(5)}_{ij}
r_{13}^{3}},\; \mbox{for $\lambda _{ij} < r_{13} \le a_{ij}$},\\
0,\; \mbox{for $r_{13} > a_{ij}$};
\end{array}
\right.
\end{array}\label{eq:dij}
\end{equation}
%%%
%%%
with $\lambda _{ij} \equiv {\displaystyle \frac{{{\left| {a_{i} -
a_{j}} \right|}}}{{2}}}$ and $a_{ij} \equiv {\displaystyle
\frac{{a_{i}+a_{j}}}{2}}$. The constants in Eq.~(\ref{eq:dij}) are
given by
%%
%%%
\begin{eqnarray*}
s_{i} &=& (n_{i} + \Gamma x_{i} ),\\ b^{(1)}_{ij}&=& 2[z_{i} n_{j}
- x_{i} s_{i} + {\frac{{a_{i}}
}{{3}}}s_{i}^{2}],\\
b^{(2)}_{ij}&=&(a_{i} - a_{j} )\left \{{\frac{{(x_{i} + x_{j}
)}}{{4}}}{\left[ {s_{i} - s_{j}}  \right]}\right.\\
&-& \left.{\frac{{(a_{i} - a_{j} )}}{{16}}}[\left( {n_{i} + \Gamma
x_{i} + n_{j} + \Gamma x_{j}}
\right)^{2} - 4n_{i} n_{j} ]\right \}, \\
b^{(3)}_{ij}& =& \left( {x_{i} - x_{j}}  \right)\left( {n_{i} -
n_{j}} \right) \nonumber \\ &+& \left( {x_{i}^{2} + x_{j}^{2}}
\right)\Gamma + \left( {a_{i} + a_{j}} \right)n_{i} n_{j} -
{\frac{{1}}{{3}}}{\left[ {a_{i} s_{i}^{2} + a_{j} s_{j}^{2}}
\right]},\\
b^{(4)}_{ij} &=& {\frac{{x_{i}} }{{a_{i}} }}s_{i} + {\frac{{x_{j}}
}{{a_{j} }}}s_{j} + n_{i} n_{j} - {\frac{{1}}{{2}}}{\left[
{s_{i}^{2} + s_{j}^{2}} \right]},\\
b^{(5)}_{ij} &=& {\frac{{s_{j}} }{{6a_{j}^{2}} }} + {\frac{{s_{i}
}}{{6a_{i}^{2}} }},
\end{eqnarray*}
where $x_{i}$ are defined as $x_{i} \equiv z_{i} + n_{i} a_{i}$
and $\Gamma$ is obtained from the solution of the following
algebraic equation
%%%
\begin{equation}
\label{eq9} \Gamma^2 = {\frac{{\pi e^{2}\beta} }{{\varepsilon}
}}{\sum\limits_{i = 1}^{n} {\rho _{i}(z_{i} + n_{i} a_{i} )^{2}}}.
\end{equation}
%%%%
The $n_{i} $ are obtained from the solution of the following set
of algebraic equations

\begin{equation}
\label{eq10}
 - \left( {z_{i} + n_{i} a_{i}}  \right)\Gamma = n_{i} + ca_{i}
{\sum\limits_{j = 1}^{n} {(z_{i} + n_{j} a_{i} )}} {\rm ,}
\end{equation}
where $c = {\displaystyle \frac{\pi}{2}}[1 - {\displaystyle
\frac{\pi}{6}}{\sum\limits_{j = 1}^{n} {\rho _{i} a_{i}^{3}
]^{-1}}} $.

Considering that $a=a_1=a_2$, $c_{ij}^{hs} (r_{13})$ is just the
direct correlation function for a hard spheres {\em binary}
mixture in the PY approximation. For particles of the same size it
is given by \cite{lebowitz64}

\begin{equation}
c_{ii}^{hs} (r_{13})= \left \{
\begin{array}{ll}
- A_{i} - B_{i} r_{13} - \delta r_{13}^{3}& \mbox{for $r_{13} < a_{i}$},\\
0, &\mbox{for $r_{13} > a_{i}$}.\\
\end{array}
\right. \label{eq:chsii}
\end{equation}
For particles of different size we have
%%%
%%%
\begin{equation}
c_{13}^{hs} (r_{13})= \left \{
\begin{array}{ll}
-A_{1}&\mbox{for $s \le \lambda_{13}$} ,\\
-A_{1} - {\frac{{\left[{\alpha x^{2} + 4\lambda_{13} \delta x^{3}
+\delta x^{4}} \right]}}{r_{13}}}&\mbox{for $\lambda_{13} < r_{13} \le a_{13}$}, \\
0  &\mbox{for $r_{13} > a_{13}$}.
\end{array}
\right.
 \label{eq:chsij}
\end{equation}
with $x\equiv r_{13}-\lambda_{13}$. The constants used in
Eqs.~(\ref{eq:chsii}) and (\ref{eq:chsij}) are given by
%%%
\begin{eqnarray}
%\begin{array}{lll}
A_1 &=& {(1-\eta_T)^{-3}}\left\{ 1 + \eta_T + \eta_T^{2}
+\frac{\pi}{6}a^3\rho_T[ 1
+2 \eta_T ]\right. \nonumber \\
&-&\left. \frac{\pi}{2}\rho_{3}(a_{3}-a)^2 \{a(1+\eta_3) +
a_{3}[1+2(\eta_1+\eta_2)]\}\right\} \nonumber\\
 &+& \frac{\pi a^3}{2}{\left(1-\eta_T\right)^{-4}}\left\{\rho_T ( 1 + \eta_T + \eta_T^{2} )\right. \\
 &-&\left. \frac{\pi}{2}\rho_{3}(\rho_1+\rho_2) (a_{3} - a)^2[\left(a + a_{3}
     \right) + aa_{3}\frac{\pi}{6}\sum_{i=1}^3 \rho_{i} a_{i}^{2}]\right\}, \nonumber \\
\alpha &=& - \pi a_{13} g_{13}(a_{13}) \sum_{i=1}^3\rho_{i}
a_ig_{ii}(a_i) , \\
\delta &=& \frac{\pi}{12}\sum_{i=1}^3\rho_{i} A_{i} ,\\
B_{1}&=& B_2=- \pi\left[ (\rho_1 +\rho_2 )a^2 g_{11}^2(a) +
\rho_{3} a_3 g_{13}^2 (a_{13} )\right],
\end{eqnarray}
with
%%%
%%%
\begin{eqnarray}
g_{11} (a)=g_{22}(a) &=& \left\{\left[ 1 + \frac{1}{2}\eta_T
\right] + \frac{3}{2}\eta _{3} a_{3}^{3} \left( a- a_{3} \right)
\right\}\left( {1 - \eta_T} \right)^{-2}, \nonumber \\
g_{13} (a_{13} )& =& \frac{[a_{3} g_{11} (a) +
ag_{33}(a_{3})]}{2a_{13}}.
\end{eqnarray}
The expressions for $A_3$, $B_{3} $ and $g_{33} \left( {a_{3}}
\right)$ are obtained by interchanging $\eta _{1}+\eta_2$,
$\rho_{1}+\rho_2$ and $a_{1} $ with $\eta _{3}$, $\rho_{3}$ and
$a_{3} $, respectively, in the expressions for $A_1$ $B_{1} $,
$g_{11}(a)$.

\subsection{The kernels expressions}
Carrying out the integrations indicated in
Eqs.~(\ref{eq:kernels1}) and (\ref{eq:kernels2}), using
Eqs.~(\ref{eq:dij}), (\ref{eq:chsii}) and (\ref{eq:chsij}), the
expressions for $K_{ij}(x,y)$ and $D_{ij}(x,y)$ are
%%%
%%%
\begin{equation}
\begin{array}{l}
D_{ij}(x,y)= \\
\left\{\begin{array}{l} b^{(1)}_{ij} k_0+ z_i z_j J_1+
b^{(2)}_{ij}M_1- b^{(3)}_{ij}M_2+b^{(4)}_{ij}M_3+
b^{(5)}_{ij} M_5,\\ \mbox{for $0 \le |x-y| \le \lambda_{ij}$},\\
(b^{(2)}_{ij}+z_{i} z_{j}) J_1 -b_{ij}^{(3)}J_2+b_{ij}^{(4)}J_3
+b_{ij}^{(5)}J_5,\\ \mbox{ for $\lambda_{ij} {\rm}  < |x - y|\le a_{ij}$},\\
0,\; \mbox{for $a_{ij}< |x - y|$}
\end{array}\right.
\end{array}
\end{equation}

\begin{equation}
-K_{ii}(x,y)=  \left\{
\begin{array}{ll}
A_{i}J_2+B_{i} J_3+\delta J_5, & \mbox{for $a_{ii}\ge |x- y|$},\\
0,&\mbox{for $a_{ii}< |x - y|$},
\end{array}\right.
\end{equation}

\begin{equation}
\begin{array}{l}
-K_{13} (x,y)= \\ \left\{
\begin{array}{ll}
A_{1}J_2 +\alpha a^{3}/3 + \delta \lambda_{13} a^{4} +
\delta a^{5}/5,& \mbox{for $|x- y|<\lambda_{13}$},\\
A_{1}J_2 + \upsilon P_3 + 4\delta \lambda_{13} P_4+ \delta P_5,\ &
\mbox{for $\lambda_{13}< |x- y|\le a_{13}$},\\
0,& \mbox{for $a_{13}<|x- y|$},
\end{array}\right.
\end{array}
\end{equation}
%%%
%%
where we use the following definitions:
\begin{eqnarray}
J_n&=&\left( {a_{ij}^{n} - |x - y |^{n}} \right)/n, \\
P_n&=&(a^{n}-(|x-y| -\lambda_{ij})^{n} )/n,\\
M_n&=&( {a_{ij}^{n} -\lambda_{ij}^{n}})/n,
\end{eqnarray}
%%%
%%%
and
%%%
\begin{equation}
k_0=(\lambda_{ij}^2 -(x-y)^2)/2.
\end{equation}

%\bibliography{donnan}

\begin{thebibliography}{10}
\bibitem{note2}
The cavity is immersed in a two component electrolyte plus one
macroion species, where the macroions are restricted to be outside
the cavity whereas
 the electrolyte can permeate it.
\end{thebibliography}
%\bibliographystyle{prsty}

\end{document}